\documentclass{ckm}                 

\usepackage{mathptmx}          

\confname{Workshop on the CKM Unitarity Triangle, IPPP Durham, April
  2003}

\title{Performance and First Physics Results of the SVT Trigger at CDF~II}

\author{I.~Vila\\
for the CDF Collaboration.}

\address{IFCA (Universidad de Cantabria - Consejo Superior de Investigaciones
Cient\'{\i}ficas), 39005  Santander - Spain }

\begin{document}

\begin{abstract}

For the first time in a hadron collider, a novel trigger processor,
the Silicon Vertex Trigger (SVT),  allows to select the long-lived
heavy flavor particles by cutting on the track impact parameter
with a precision similar to that of the offline reconstruction. Triggering
on displaced tracks has enriched the B-physics program by enhancing the B
yields of the lepton-based triggers and opened up full hadronic
triggering at CDF. After a first commissioning period, the SVT is
fully operational, performing very closely to its design capabilities. System
performance and first physics results based on SVT selected data
samples are presented.


\end{abstract}


\maketitle

\section{Introduction}

The upgraded Collider Detector at Fermilab (CDF) at the
$p\bar{p}$  Tevatron collider  is back in operation since March
2001. A detailed description of the upgraded detector can be found
elsewhere\cite{TDR}. New vertexing, triggering and particle
identification capabilities have been added enhancing the potentialities of the
B physics program\cite{Rolf}.

There are several motivations for pursuing B physics at  a
$p\bar{p}$ collider at $\sqrt{s}=1.96~TeV$. The first reason is
the three orders of magnitude higher production rate of B hadrons
with respect to the conventional B experiments in $e^+e^-$
colliders at the $\Upsilon(4S)$ or at the $Z^0$ pole. This huge B
production cross section allows searches for the very rare B
decays and performing precise B physics measurements. Another important
advantage is the fact that all the B hadron species are produced,
in contrast with the $e^+e^-$ machines at the $\Upsilon(4S)$
experiments where only the $B^+$ and $B^0_d$ are created.

However, QCD backgrounds in a hadronic experiment are also three
orders of magnitude higher, therefore dedicated triggers for
suppressing the background are needed. For the first time in a
hadron detector, a novel  Silicon Vertex Trigger (SVT) is being
used. This new level-2 trigger processor  allows to select the
long-lived heavy flavor particles by cutting on the track impact
parameter with a precision similar to that achieved by the full
event offline reconstruction.

In the following sections, a brief description of the SVT, its
performance and first physics results are given.

\section{SVT Overview}


\subsection{CDF trigger system and SVT functionalities}

 The CDF trigger system has also been upgraded to accommodate the
 new bunch spacing (396~ns) and larger luminosity ($\sim
 10^{32}~cm^2s^{-1}$). The CDF Trigger is a three level pipelined
 and buffered system. The first level is a deadtimeless pipelined synchronous 
 system with
 a present accept rate of about 20~kHz and a pipeline depth of 5.5~$\mu s$. 
 The first
 level trigger decision is made using the transverse energy of the calorimeter,
  tracks in the 
central drift chamber and track segments in the muon system. The
 second level trigger is an asynchronous system with a latency of
 $\sim$ 30~$\mu$s and a current accept rate of ~300~Hz.
It is structured as a two stage pipeline with data buffering at
the input of each stage. The first stage is based on dedicated
hardware processors which assemble information from a particular
section of the detector (Calorimeter Clusters, Silicon tracks,
etc). The second stage consists of programmable processors (DEC
Alpha processors) operating on lists of objects generated by the
first stage. Each of the L2 stages is expected to take
approximately half of the total level-2 latency. The SVT works inside
the first stage of this trigger level. The final selection is
 done by the level-3 event filter, a farm of about 250 Linux PCs which
performs the full event reconstruction, reducing
typically to $\sim$~50~Hz the writing rate to permanent
storage.

The SVT  is a data driven trigger processor at the level-2 of the
CDF trigger system, dedicated to the 2-D reconstruction of charged
particle trajectories in a plane perpendicular to the beam
line~\cite{SVT}. The SVT inputs are the sparsified list of channel
numbers from four layers of the Silicon Vertex detector (SVX~II)
~\cite{SVXII} with its corresponding pulse amplitudes, and the
tracks reconstructed in the central drift chamber (COT)~\cite{COT}
by the new level-1 trigger processor, the eXtremely Fast Tracker
(XFT)~\cite{XFT}. With a latency of only $1.9~\mu s$, the XFT
finds and fits COT tracks of transversal momentum larger than
$1.5~GeV$ with  a tracking efficiency of 96\%. The SVT output is a
list of reconstructed tracks with a core resolution of the track
parameters very similar to that of the offline track reconstruction.

The SVT performs SVX~II hit finding, pattern recognition and track
fitting in about $15~\mu s$ within the time window of the level-2
first stage latency. The key design features to accomplish these
tasks in such a short time interval are a \emph{ highly parallel
design} (architecture and tasking), a \emph{pipelined
architecture}, a \emph{custom VLSI pattern recognition} and
\emph{linearized fitting} of the particle candidate trajectories.

\subsection{System architecture and data processing}

\begin{figure}
\hbox to\hsize{\hss
\includegraphics[width=\hsize,clip=]{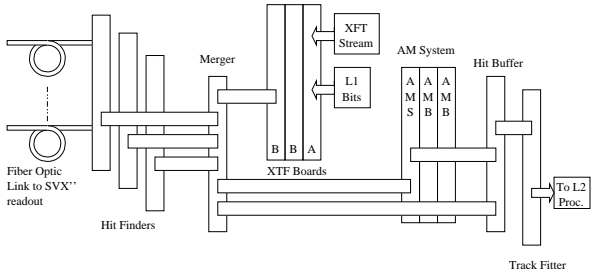}
\hss} \caption{Board organization of an  SVT sector ($1/12$ of the
whole system) .}
 \label{fig:SVTarch}
\end{figure}

The SVT has been implemented on custom design VME 9U boards,
organized in 12 identical subsystems (sectors) running
independently in parallel. This segmentation maps the SVX~II
geometry, which is divided into 12 identical wedges along the
azimuthal angle. The main functional blocks of each SVT sector are
the Hit Finders, the Associative Memory system, the Hit Buffer and
the Track Fitter, see Fig.~\ref{fig:SVTarch}.

Raw SVX~II data flow from the front-end to the Hit Finder boards
that find clusters of strips with a significant energy deposit and
compute the coordinate of the centroids (hits). The curvature and
the azimuthal angle of COT tracks from the XFT are received,
fanned out to the 12 SVT sectors and fed both to the Associative
Memory and to the Hit Buffer boards together with the SVX~II hits.
The Associative Memory boards find coincidences between the SVX~II
hits and a set of predefined patterns of particle trajectories. To
reduce the amount of required memory to store those trajectory
templates, the pattern recognition process is performed at a
coarser resolution than the full available detector resolution,
choosing a SuperStrip size of 250~$\mu m$ for the silicon layers
and $5^{o}$ for the azimuthal XFT track angle.

The output of the Associative Memory system is the list of track
candidates (roads),which are sent to the Hit Buffer. Each Hit
Buffer stores all hits and tracks in a sector for each event, then
retrieves the hits and the XFT tracks belonging to
each road and sends them to the Track Fitter boards that perform
quality cuts and estimate track parameters using the full
available spatial resolution in a linearized fit.

A set of Merger boards, each providing 4-way fan-in and 2 way
fan-out, ties together all boards in a chain that starts from 144
optical links from the SVX~II, 1~Gbit/s each, and ends in a single data
path, a 0.7 Mbit/s LVDS cable, to the CDF level-2 processor.

\subsection{SVT Performance}





     %







The SVT performance has been evaluated with data samples selected
by triggers which do not require SVT information, like the
"dimuon" or "minimum bias" trigger paths, to avoid any possible
bias on the results.

The precise determination of the transverse impact parameter of
the SVT tracks requires the previous knowledge of the actual beam
line position, the real SVX~II geometry and the relative position
between the central drift tracker chamber and the SVX~II.
Originally, the baseline plan was to steer the Tevatron beam to
its nominal position using the SVT as a beam position monitoring
device. It turns out to be more convenient to do a
real-time determination of the beam position (computed every
30~s), using the SVT built-in data monitoring and diagnosis
capabilities, and correct the fitted track parameters in an event
by event basis, avoiding the need of  modifying the beam orbit.

\begin{figure}
\hbox to \hsize{\hss
\includegraphics[width=200pt p,clip=]{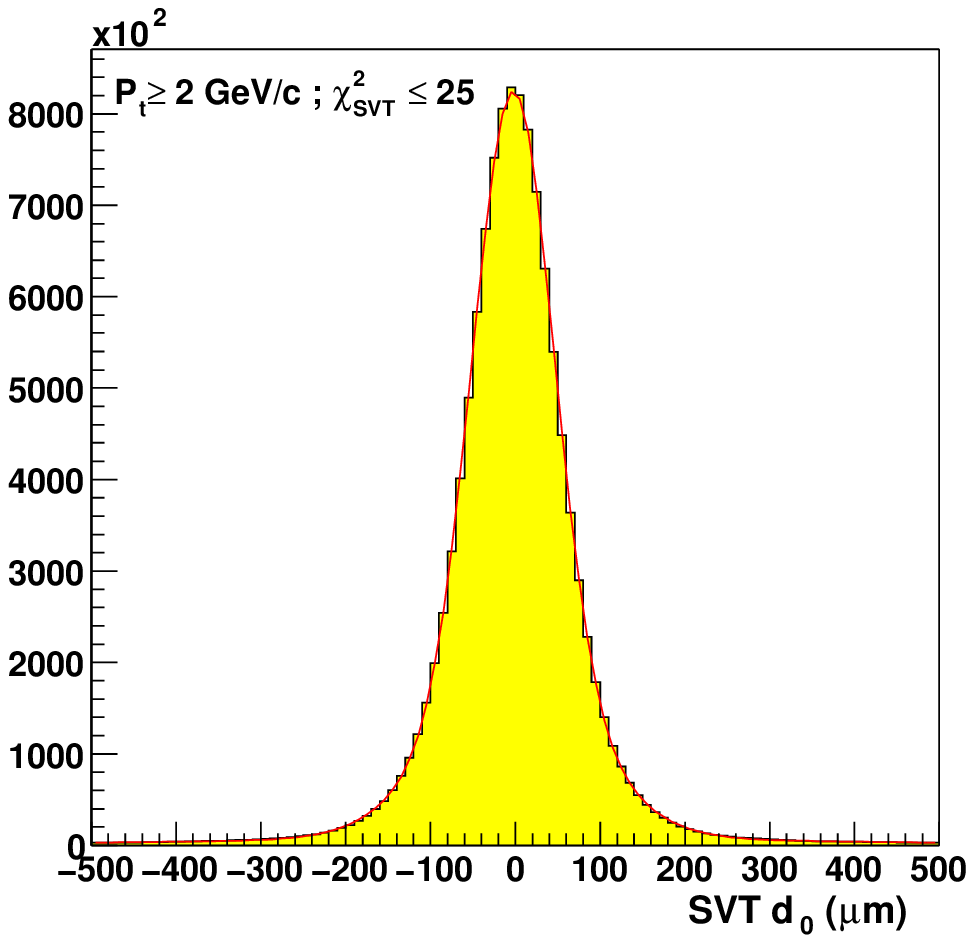}
\hss} \caption{SVT tracks impact parameter distribution.}
 \label{fig:SVTd0}
\end{figure}

The impact parameter distribution of the SVT tracks corrected by
SVX~II alignment and the beam line position is shown in
Fig.~\ref{fig:SVTd0}. The distribution width $\sim 50~\mu m$
results from the convolution of 35~$\mu m$ SVT resolution and the
beam spot with a 33~$\mu m$ width.

The overall SVT tracking efficiency is about 80\%. This
efficiency is defined as the ratio between the number of SVT
tracks and the number of offline tracks matched to a XFT track in
the SVT fiducial volume. No tracking efficiency dependence on
the transverse momentum of the tracks is observed.

\begin{figure}
\hbox to\hsize{\hss
\includegraphics[width=200pt,clip=]{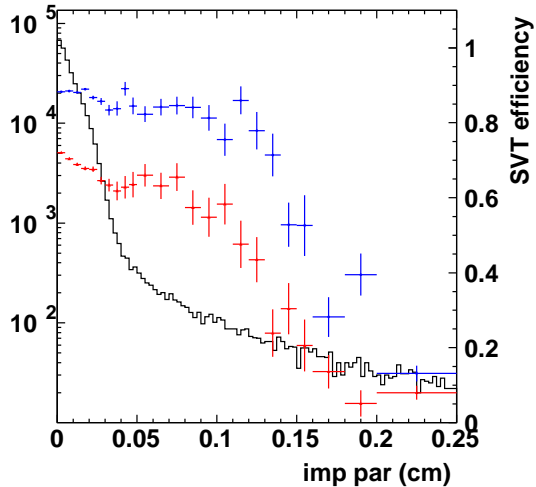}
\hss} \caption{SVT tracking efficiency dependence on track impact
parameter. The distribution of tracks is shown in black, the SVT
efficiency with the initial (new) road patterns in clearer (darker).}
 \label{fig:Effd0}
\end{figure}

 In the case of the efficiency dependence on
the impact parameter of the tracks, shown in Fig.~\ref{fig:Effd0}, we
observe a sharp drop in the efficiency for impact parameter
values beyond 0.1~cm. This effect is due to the finite number  of
predefined road patterns that can be stored in the Associative
Memory boards.A new improved pattern set, which accounts for a
larger  than expected shift between the SVX~II and the COT, and
the beam line offset, has increased by $\sim 10\%$ the initial SVT
overall tracking  efficiency. The other significant efficiency improvement
comes from the SVX~II acceptance recovery.


There is always the possibility that random combinations of SVX~II
hits which do not correspond to real tracks pass the SVT track
quality cuts producing fake tracks. The overall SVT track purity
is about 80\%; the purity is close to 100\% for primary tracks and
gets lower for high impact parameter tracks.

\section{Initial Physics Signals}

Conventionally, triggering  on dimuons coming from $J/\psi$
or semileptonic b decays was the only way to do heavy
flavor physics at hadron colliders. The SVT has opened the new way
of fully hadronic decay triggering, and expanded the
semileptonic trigger. The initial experience with the SVT has been
very successful, allowing for competitive results even with a
modest amount of integrated luminosity.



%
\subsection{Semileptonic trigger}

The enhanced semileptonic trigger selects events with a lepton of
transverse momentum  $p_T > 4~GeV$, and a displaced track with an
impact parameter $100~\mu m < d_0 < 1~mm$ with $p_T > 2~GeV$. The
additional displaced track requirement allows for a lower lepton $p_T$ cut,
increasing the collected sample size. For a total integrated
luminosity of about $70 pb^{-1}$ there are more than half a
million recorded inclusive semileptonic decays, a yield five
times larger than the Run~I equivalent semileptonic trigger. For
the same luminosity, the number of partially reconstructed
semileptonic charmed b decays, like $B \rightarrow lD^0X$ ($D^0
\rightarrow K \pi$), $B \rightarrow lD^*X$ ($D^* \rightarrow
D^0\pi$) and $B \rightarrow lD^+X$ ($D^+ \rightarrow K \pi\pi$) is
about 10000, 1500 and 5000 events respectively. These samples are
used for flavor tagging optimization and measurements of $B_d$ mixing 
and life times.


\subsection{Fully hadronic trigger}

The hadronic trigger selects events with at least two displaced
tracks of opposite charge with transverse momentum $p_T
> 2~GeV$ and $\sum p_T > 5.5~GeV$. It comes in two flavors for
two different event topologies: the two-body decay trigger with an
impact parameter cut of $100~\mu m < d_0 < 1~mm$ and large opening
angle between trigger tracks, and the multibody decay trigger with
tighter impact parameter cut ($120~\mu m < d_0 < 1~mm$) and
smaller opening angle. Tracks originating from tertiary decay vertexes in a
multibody decay will have on average larger impact parameters than tracks from the
secondary decay vertexes in a two-body decay. 

Originally, the hadronic trigger was designed for collecting fully
hadronic b decays. It turns out to be also very efficient
collecting charm hadronic decays.

In the following sections, we will give a brief summary of the
main initial charm and b physics results based on the SVT hadronic
trigger.

\subsubsection{Charm physics results}


The measurement of the prompt D meson cross section is of
theoretical interest to clarify the larger than expected beauty
cross section compared to the next-to-leading order QCD
calculations. A preliminary measurement of the D meson cross
sections was done with a total integrated luminosity of $5.7 ~pb^{-1}$. The production cross
sections were found to be larger than the corresponding bottom
meson cross sections: $4.3\pm 0.1 \pm 0.7 \mu b$ for $D^+$  and
$9.3\pm 0.1 \pm 1.1 \mu b$ for $D^0$.

\begin{figure}
\hbox to\hsize{\hss
\includegraphics[width=200pt,clip=]{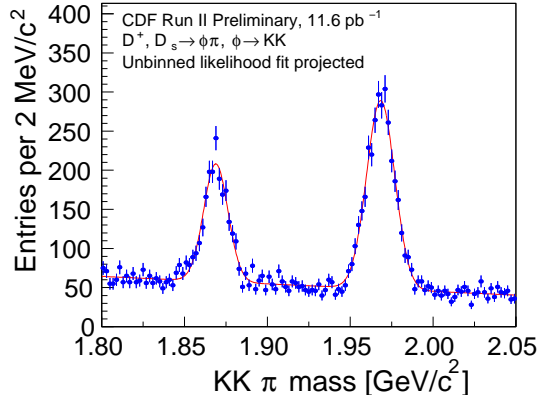}
\hss} \caption{Invariant mass distribution for $D^+_s$, $D^+$
$\rightarrow \phi \pi$ signals with superimposed fit.}
 \label{fig:DeltaM}
\end{figure}

The measurement of the $m_{D_s^+}- m_{D^+}$ mass difference
provides a test for the Heavy Quark Effective Theory and QCD. The
analysis required the precise calibration of the momentum scale of
the detector using a sample of 50000 $J/\Psi \rightarrow \mu \mu$
decays. The calibration was applied to the decays $D_s^+, D^+
\rightarrow \phi \pi^+$ and the mass difference value found was
$m_{D_s^+}- m_{D^+} = 99.41 \pm~0.38 (stat.) \pm~0.21(syst.)
~MeV/c^2$~\cite{DeltaM}. Figure~\ref{fig:DeltaM} shows the invariant mass
distribution of the two signals.

Observation of large CP violation in the charm system would be an
evidence of new physics beyond the Standard Model. In a sample of
$65~\pm 4~pb^{-1}$, the Cabibbo-suppressed decays to CP
eigenstates $D^0 \rightarrow K^{+}K^{-}$ and $D^0 \rightarrow
\pi^{+}\pi^{-}$ were reconstructed. All the $D^0$ were selected
originating from a $D^{*+}$ decay $D^{*+} \rightarrow D^0 \pi^+$
which provides a very pure and flavor tagged sample. The direct CP
asymmetries $A = \frac{\Gamma(\bar D \rightarrow \bar f)-\Gamma(D \rightarrow
f)} {\Gamma(\bar D \rightarrow \bar f)+\Gamma(D \rightarrow f)}$ were found to be $2.0~\pm~1.7(stat.)~\pm~0.6(syst.)\%$
for $D^0 \rightarrow K^{+}K^{-}$ and
$3.0~\pm~1.9(stat.)~\pm~0.6(syst.)\%$ for $D^0 \rightarrow
\pi^{+}\pi^{-}$ decay modes.


\subsubsection{Beauty physics}

\begin{figure}
\hbox to\hsize{\hss
\includegraphics[width=200pt,clip=]{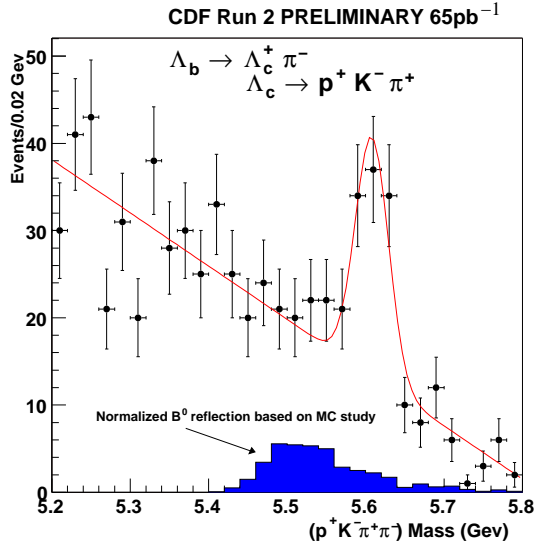}
\hss} \caption{$\Lambda_b$ signal reconstructed in the $\Lambda_b
\rightarrow \Lambda_c^+ \pi^-$ channel.}
 \label{fig:lambda}
\end{figure}

Initially purely hadronic b decays have been established, and currently
we are in the process of understanding the signal yields. In
particular, events  for the $B_s$ mixing golden channel $Bs
\rightarrow D_s \pi$ has been observed~\cite{Donatella}. Fully reconstructed 
$\Lambda_b$'s has been
collected, using the $\Lambda_b \rightarrow \Lambda_c^+ \pi^-$
decay with $\Lambda_c \rightarrow p^+K^-\pi^+$. In in Fig.~\ref{fig:lambda}.
the $\Lambda_b$ invariant mass distribution is shown. 

\begin{figure}
\hbox to\hsize{\hss
\includegraphics[width=200pt,clip=]{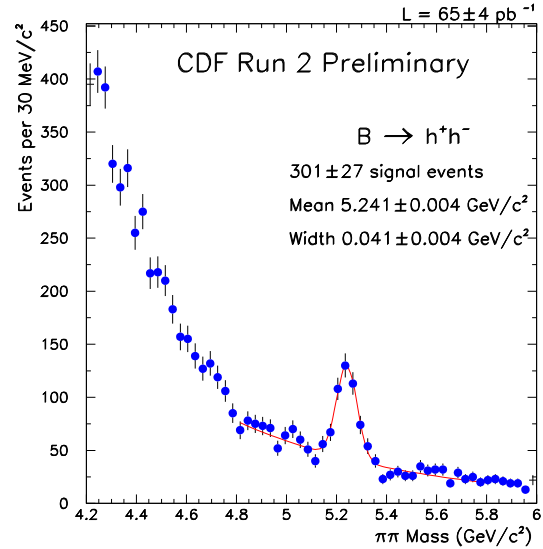}
\hss} \caption{Invariant mass distribution for $B \rightarrow h^+ h^-$
signals with superimposed fit}
 \label{fig:btohh}
\end{figure}

The charmless purely hadronic two-body decay of neutral B mesons
has been reconstructed. The invariant mass distribution of two
tracks of opposite charge, assuming the pion hypothesis for each
track, is shown in Fig.\ref{fig:btohh}. The mass peak is a mixture
of $B^0,B_s \rightarrow \pi\pi, K\pi, KK$ decays. The contribution
of each decay mode has been disentangled with discriminating cuts
based on track dE/dx and kinematics. The physics interests of this
measurement are, among others, the first evidence of charmless $B_s$
decays, direct CP asymmetry measurements in $B^0 \rightarrow
K^+\pi^-$ and  CP asymmetry in $B^0 \rightarrow \pi\pi.$

\begin{thebibliography}{9}
\bibitem{TDR}The CDF-II Colaboration, FERMILAB-PUB-96-390-E, Nov 1996.
\bibitem{Rolf}
R.G.C~Oldeman, \emph{Performance of CDF for B physics}, these
proceedings.
\bibitem{SVT} W. Ashmanskas et al.,IEEE
Trans.Nucl.Sci. 49, 1177 (2002).
\bibitem{SVXII} A. Sill [CDF Collaboration], Nucl. Instrum. and
Meth. A 447, 1 (2000).
\bibitem{COT} K.T. Pitts[CDF Collaboration], Nucl. Phys. Proc.
Suppl. 61B, 230 (1998).
\bibitem{XFT} E. J. Thomson et al., IEEE Trans. Nucl. Sci. 49,
1063 (2002).
\bibitem{DeltaM} D. Acosta et al.[CDF-II Collaboration], Measurement 
of the Mass Difference $m(D_s^+)- m(D^+)$ at CDF II. Submitted to Phys. Rev. D.
\bibitem{Donatella}
D~Lucchesi, $B_s$ Physics and Prospects at the Tevatron, these
proceedings.


\end{thebibliography}
\end{document}